\documentclass[twocolumn,showpacs,preprintnumbers,amsmath,amssymb,prb]{revtex4}
\usepackage{graphicx}% Include figure files
\usepackage{dcolumn}% Align table columns on decimal point
\usepackage{bm}% bold math
\usepackage{enumerate}

\begin{document}

\preprint{APS/123-QED}

\title{Identification of microscopic spin-polarization coupling in the ferroelectric phase of %
a magnetoelectric multiferroic CuFe$_{1-x}$Al$_x$O$_2$}

\author{Taro Nakajima}
\email{E-mail address: nakajima@nsmsmac4.ph.kagu.tus.ac.jp}
\author{Setsuo Mitsuda}
\affiliation{Department of Physics, Faculty of Science, Tokyo University of Science, Tokyo 162-8601, Japan}%
\author{Toshiya Inami}
\affiliation{Japan Atomic Energy Agency, Koto, Sayo-cho, Sayo-gun, Hyogo, 679-5148, Japan}%
\author{Noriki Terada}
\affiliation{ICYS, National Institute for Materials Science, Tsukuba, Ibaraki 305-0044, Japan}%
\author{Hiroyuki Ohsumi}
\affiliation{RIKEN SPring-8 Center, 1-1-1 Kouto, Sayo-cho, Sayo-gun, Hyogo 679-5148, Japan}%
\author{Karel Prokes}
\author{Andrei Podlesnyak}
\affiliation{Hahn-Meitner Institute, SF-2, Glienicker Str. 100, 14109 Berlin, Germany }%

\begin{abstract}
We have performed synchrotron radiation X-ray and neutron diffraction measurements %
on magnetoelectric multiferroic CuFe$_{1-x}$Al$_x$O$_2$ ($x=0.0155$), which %
has a proper helical magnetic structure with incommensurate propagation wave vector in the %
ferroelectric phase. %
The present measurements revealed that the ferroelectric phase is accompanied by %
lattice modulation with a wave number $2q$, where $q$ is the magnetic modulation wave number. %
We have calculated the Fourier spectrum of the spatial modulations in the local electric polarization %
using a microscopic model proposed by Arima [T. Arima, J. Phys. Soc. Jpn. {\bf 76}, 073702 (2007)]. %
Comparing the experimental results with the calculation results, we found that the origin of the $2q$-lattice modulation is %
not conventional magnetostriction %
but the variation in the metal-ligand hybridization between the magnetic Fe$^{3+}$ ions and ligand O$^{2-}$ ions. %
Combining the present results with the results of a previous polarized neutron diffraction study [Nakajima \textit{et al.}, Phys. Rev. B {\bf 77} 052401 (2008)], %
we conclude that the microscopic origin of the ferroelectricity in CuFe$_{1-x}$Al$_x$O$_2$ is %
the variation in the metal-ligand hybridization with spin-orbit coupling. %
\end{abstract}

\maketitle

\section{INTRODUCTION}

Magnetically induced ferroelectricity or electric control of magnetic ordering has been intensively %
investigated since a colossal magneto-electric (ME) effect was found in TbMnO$_3$.\cite{Kimura_nature} %
Recent experimental studies have discovered a variety of ferroelectric magnetic compounds, %
which might be suitable for use in advanced ME devices. %
These materials are often termed ME multiferroics. %
In order to develop practical device applications, it is essential to determine the microscopic mechanism %
of spin-polarization coupling in these systems. %
So far, there have been two different microscopic models %
for magnetically induced ferroelectricity. %
One is the spin-current model,\cite{Katsura_PRL_2005} %
which predicts that two noncollinearly aligned neighboring spins $\mbox{\boldmath $S$}_i$ and $\mbox{\boldmath $S$}_{i+1}$
generate a local electric dipole moment $\mbox{\boldmath $p$}$ given by %
$\mbox{\boldmath $p$}\propto\mbox{\boldmath $e$}_{i,i+1}\times (\mbox{\boldmath $S$}_i \times \mbox{\boldmath $S$}_{i+1})$, %
where $\mbox{\boldmath $e$}_{i,i+1}$ is the unit vector connecting the two spins. %
This formula predicts macroscopic uniform electric polarization in a magnetic structure with cycloidal spin-components, %
and shows excellent agreement with the experimentally determined relationships between magnetic structures and electric polarization %
in various transition metal oxides, such as TbMnO$_3$,\cite{Kenzelmann_PRL_Spiral} %
Tb$_{1-x}$Dy$_x$MnO$_3$,\cite{Arima_PRL_Spiral} %
Ni$_3$V$_2$O$_8$,\cite{PRL_Ni3V2O8} %
MnWO$_4$,\cite{PRL_MnWO4} %
and CoCr$_2$O$_4$.\cite{Yamasaki_PRL_conical} %
The other microscopic model is the magnetostriction model, which predicts ferroelectricity in collinear-commensurate magnetic structures. %
In this case, the local electric dipole moment is given by %
$C(\mbox{\boldmath $r$})(\mbox{\boldmath $S$}_i\cdot \mbox{\boldmath $S$}_{i+1})$, where $C(\mbox{\boldmath $r$})$ is a constant dependent on %
the local crystal structure and the exchange interactions. % 
The spontaneous electric polarization observed in the collinear-commensurate magnetic orderings in orthorhombic %
HoMnO$_3$\cite{HoMnO3} %
%and TmMnO$_3$\cite{TmMnO3} 
may be attributed to this model. %
However, there are several ME multiferroic materials whose ferroelectricity cannot be explained by either of these models,\cite{RFMO} %
for example, delafossite multiferroic CuFeO$_2$.\cite{Kimura_CuFeO2,SpinNoncollinearlity,CFAO_Helicity} %
Hence, identification of microscopic spin-polarization coupling in these systems paves the way for %
a new design of multiferroic materials. %

\begin{figure}[t]
\begin{center}
	\includegraphics[keepaspectratio,clip,width=7.8cm]{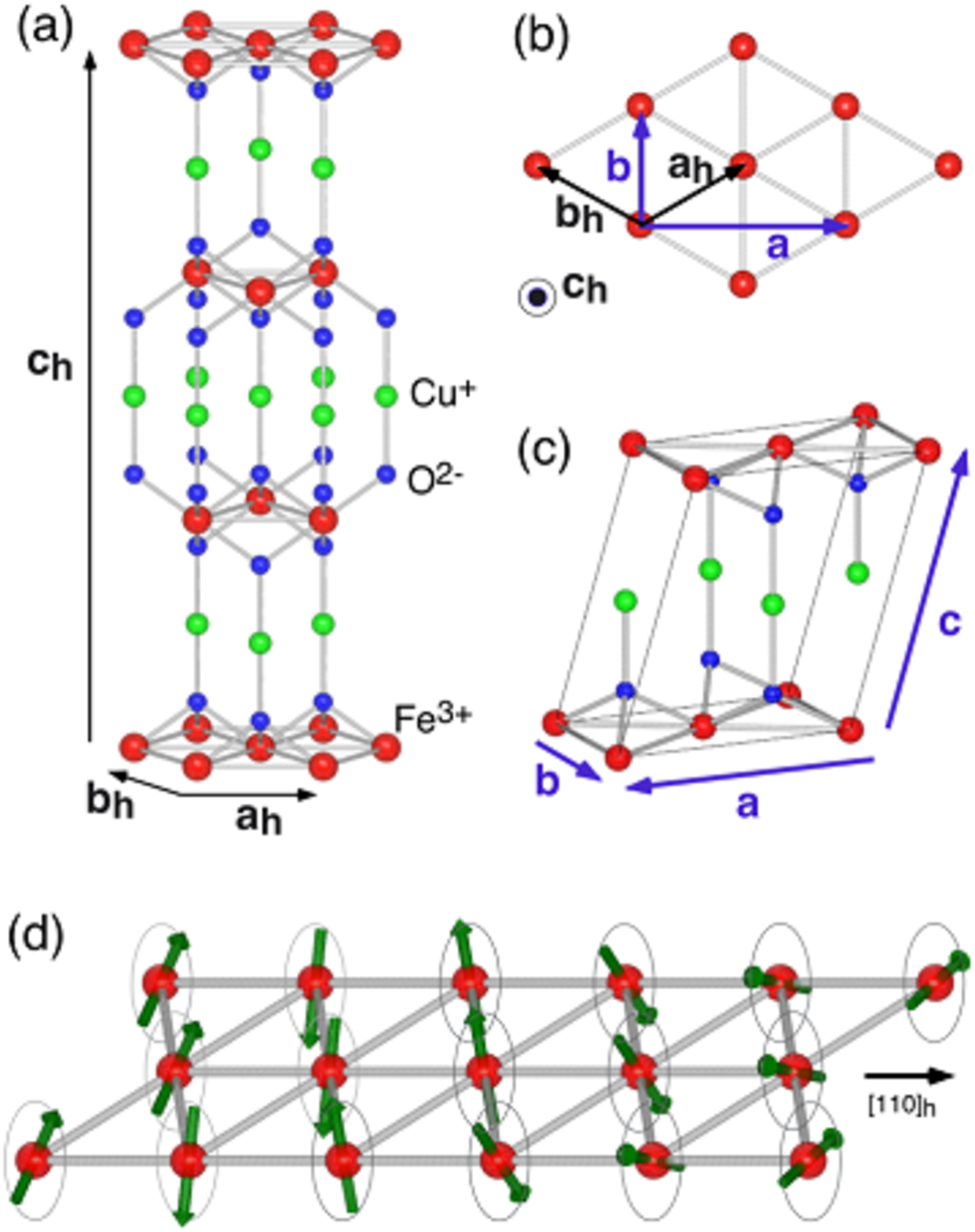}
	\caption{(Color online) (a) Crystal structure of CuFeO$_2$ with the hexagonal basis ($\mbox{\boldmath $a$}_h ,\mbox{\boldmath $b$}_h, \mbox{\boldmath $c$}_h$). (b)-(c) Relationship between the hexagonal basis %
	and the monoclinic basis ($\mbox{\boldmath $a$} ,\mbox{\boldmath $b$}, \mbox{\boldmath $c$}$). %
	(d) Schematic drawing of the proper helical magnetic structure in CuFe$_{1-x}$Al$_x$O$_2$. }
	\label{structure_phasediagram}
\end{center}
\end{figure}

\begin{figure}[t]
\begin{center}
	\includegraphics[keepaspectratio,clip,width=7.8cm]{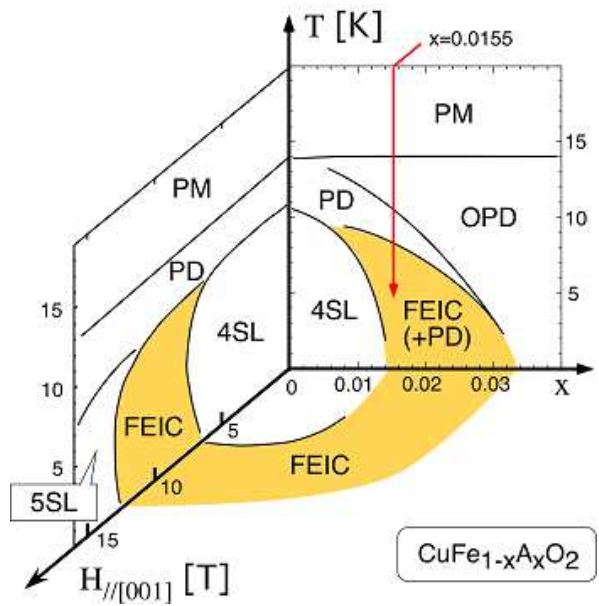}
	\caption{(Color online) Schematic drawing of $x$-$H$-$T$ phase diagram of CuFe$_{1-x}$Al$_x$O$_2$.}
	\label{x-T-H_PhaseDiagram}
\end{center}
\end{figure}

The crystal structure of CuFeO$_2$ is shown in Figs. \ref{structure_phasediagram}(a)-(c). %
Owing to the geometrical frustration in the triangular lattice planes of the magnetic Fe$^{3+}$ ions, %
CuFeO$_2$ exhibits various magnetically ordered phases.\cite{Mitsuda_1991,Mitsuda_2000,Petrenko_2000} %
The ground state of CuFeO$_2$ is a collinear-commensurate 4-sublattice (4SL) antiferromagnetic state. %
A spontaneous electric polarization %
emerging in the direction perpendicular to the hexagonal $c$ axis was discovered in the first field-induced phase.\cite{Kimura_CuFeO2} %
Subsequent studies\cite{Kanetsuki_JPCM,Seki_PRB_2007} %
revealed that the ferroelectric phase is stabilized even under %
zero field by substituting a small amount of nonmagnetic Al$^{3+}$ ions for magnetic Fe$^{3+}$ ions, as seen in Fig. \ref{x-T-H_PhaseDiagram}. %
%
%By the recent neutron diffraction measurements under applied field, %
%the magnetic structure of the ferroelectric phase has been elucidated to be an antiferromagnetically stacked proper helical magnetic structure %
%with incommensurate wave vector.\cite{SpinNoncollinearlity}  %
Recent neutron diffraction measurements in applied field revealed that %
the ferroelectric phase magnetic structure is an antiferromagnetically %
stacked proper helical magnetic structure.\cite{SpinNoncollinearlity}  %
The wave vector is incommensurate. %
The helical axis is parallel to the hexagonal [110] direction, as shown in Fig. \ref{structure_phasediagram}(d). %
Hereafter, we refer to the ferroelectric phase as the ferroelectric incommensurate (FEIC) phase. %
Since neither the spin-current model nor the magnetostriction model is able to explain the ferroelectricity in proper helical magnetic %
ordering, we anticipate that another type of spin-polarization coupling is realized in CuFe$_{1-x}$Al$_x$O$_2$. %

Recently, Arima presented a theoretical consideration on the ferroelectricity in CuFe$_{1-x}$Al$_x$O$_2$ %
suggesting that the variation in the metal-ligand hybridization with spin-orbit coupling is %
relevant to the ferroelectricity in this system. %
Applying a microscopic theory derived by Jia \textit{et al.}\cite{Jia1,Jia2} to a cluster model with a proper helical spin arrangement, %
Arima has predicted the following intrinsic features of the ferroelectricity in this system: %

Feature (i): the direction of the uniform polarization should be parallel to the helical axis of the proper helical magnetic ordering. %

Feature (ii): the spin helicity (i.e., right- or left-handed helical arrangement of spins) should correspond to the polarity of %
the uniform polarization. %

Feature (iii): There must be spatial modulations with wave numbers of $2q$ and $4q$ (where $q$ is the magnetic modulation wave number) %
in the helical-axis components of the local electric polarization vectors. %

The results of recent polarized neutron diffraction measurements on CuFe$_{1-x}$Al$_x$O$_2$ ($x=0.02$) under an applied %
electric field show excellent agreement with features (i) and (ii).\cite{CFAO_Helicity} %
However, the existence of spatial modulations of the local polarization (feature (iii)) has not yet been confirmed. %
Spatial modulations of the local polarization must result in lattice modulations, which can be observed by X-ray diffraction %
measurements. %
The existence of incommensurate lattice modulations has been reported in the field-induced FEIC phase of undiluted CuFeO$_2$ by %
Terada \textit{et al.}\cite{Terada_LatticeModulation} and Ye \textit{et al.}\cite{Ye_CuFeO2} %
However, application of a magnetic field readily induces slight distortion of the magnetic structure %
as well as the uniform magnetization components. %
Actually, higher harmonic magnetic reflections have been detected in the field-induced FEIC phase of CuFeO$_2$.\cite{Mitsuda_2000} %
As discussed by Terada \textit{et al.} in Ref.\onlinecite{Terada_LatticeModulation}, this situation can easily cause additional lattice modulations through magnetostriction. %
In order to elucidate the existence of the lattice modulations predicted by the metal-ligand hybridization model (Feature (iii)), %
it is essential to remove the contribution of these field-induced lattice modulations. %

In the present study, %
we have performed synchrotron radiation X-ray measurements using a CuFe$_{1-x}$Al$_x$O$_2$ ($x=0.0155$) %
sample, which exhibits a FEIC phase under zero field. %
The present measurements reveal %
the existence of $2q$-lattice modulations in the zero-field FEIC phase. %
We have also investigated the magnetic-field dependence of the $2q$ lattice modulations by %
X-ray and neutron diffraction measurements under a field applied along the helical axis. %
In order to identify the origin of the observed $2q$-lattice modulation, %
we calculate the Fourier spectrum of the spatial modulations in the helical-axis component of the %
local electric polarization using %
the microscopic model presented by Arima.\cite{Arima_Symmetry} %
This calculation reveals that the lattice modulation does not originate from a conventional magnetostriction %
but rather from the variation in the metal-ligand hybridization between the magnetic Fe$^{3+}$ ions and the ligand O$^{2-}$ ions. %
As a result, we conclude that the microscopic mechanism of the ferroelectricity in CuFe$_{1-x}$Al$_x$O$_2$ is %
the variation in the metal-ligand hybridization with spin-orbit coupling. %

\begin{figure*}[t]
\begin{center}
	\includegraphics[keepaspectratio,clip,width=17cm]{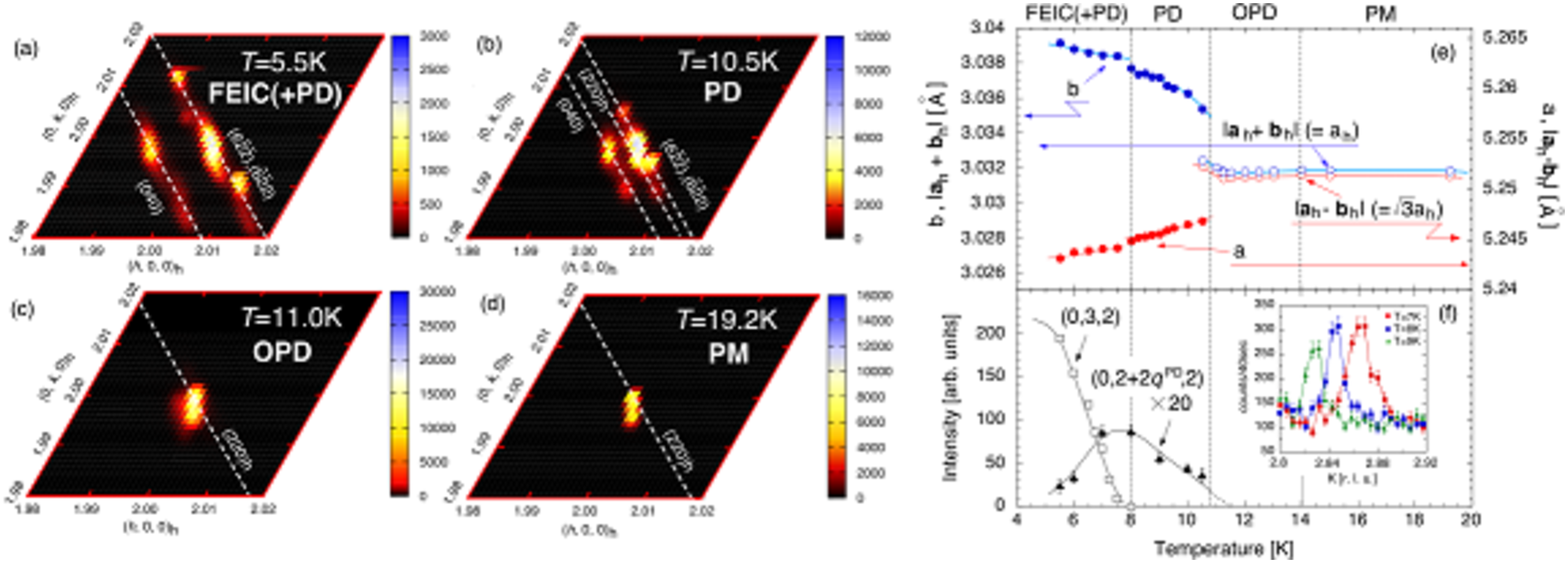}
	\caption{The X-ray diffraction intensity maps obtained by the $(H,K,0)_h$ reciprocal lattice scans around the reciprocal lattice position of $(2,2,0)_h$ at (a) $T=5.5$ K, (b) $T=10.5$ K, %
	(c) $T=11.0$ K and (d) $T=19.2$ K. These intensity maps were measured in a warming process. %
	(e) Temperature variations of the lattice constants $a$ and $b$ in the monoclinic notation, which were measured on heating. %
	(f) Temperature variations of the intensities of $(0,3,2)$ and $(0,2+2q^{\rm PD},2)$ reflections. The inset shows the diffraction %
	profiles of the $(0,2+2q^{\rm PD},2)$ reflections at typical temperatures. The data of (0, 2 + 2qPD, 2) are  %
	magnified by a factor of 20. }
	\label{BL46XU_results}
\end{center}
\end{figure*}

\section{EXPERIMENTAL DETAILS}
A single crystal of CuFe$_{1-x}$Al$_x$O$_2$ with $x=0.0155$ was %
prepared by the floating zone technique \cite{Zhao_FZ}, and was cut into a disk shape. %
The Al-concentration was determined by chemical analysis. %
We performed X-ray diffraction measurements in zero field at the beamline BL46XU in SPring-8. %
The energies of the incident X-ray beams were tuned to 12 keV. %
The sample was mounted in a closed-cycle $^4$He-refrigerator. % 
The data obtained in this measurement are presented in sec. III. A.
%The data presented in sec. III. A were obtained in this measurements. %

We have also performed X-ray diffraction measurements in zero field and in fields applied along the hexagonal $[110]$ %
(monoclinic $b$ axis) at the beamline BL22XU in SPring-8. %
The sample was mounted in a horizontal-field cryomagnet whose maximum field is 6 T; %
the hexagonal $(hhl)$ plane was used as the scattering plane. %
The data obtained in these measurements are presented in sec. III, B and C. %
%The data presented in sec. III. B and C were obtained in this measurements. %

Neutron diffraction measurements in a field applied along the hexagonal $[110]$ axis were performed at the two-axis neutron diffractometer E4 %
installed at the Berlin Neutron Scattering Center in Hahn-Meitner Institute.%
The hexagonal ($hhl$) plane was selected as the scattering plane. %
Incident neutrons with wave numbers of 2.44 \AA were obtained by a pyrolytic graphite (002) monochromator. %
An external magnetic field directed along the hexagonal $[110]$ direction was provided by the horizontal field cryomagnet, %
HM-2, whose maximum field is 4 T. %

As described later, CuFe$_{1-x}$Al$_x$O$_2$ with $x=0.0155$ exhibits a symmetry-lowering structural transition from a rhombohedral structure %
to a structure with lower symmetry (probably monoclinic symmetry, but possibly even lower symmetry). Taking into %
account the monoclinic lattice distortions found in the magnetically %
ordered phases of undiluted CuFeO$_2$,\cite{Terada_14.5T,Terada_CuFeO2_Xray,Ye_CuFeO2} %
it is reasonable to employ a monoclinic basis in addition to the conventional hexagonal basis. %
The definitions of these bases are shown in Figs. \ref{structure_phasediagram}(a)-(c). %
We mainly employed the monoclinic notation. %
To distinguish between the two bases, the subscript 'h' has been added to the hexagonal %
notation when referring to modulation wave numbers and reciprocal lattice indices. %

\section{EXPERIMENTAL RESULTS}
\subsection{Characterization of the crystal structures in the magnetically ordered phases}

Before discussing the lattice modulations in the ferroelectric phase, we present the symmetry-lowering structural transition %
corresponding to the magnetic phase transitions in CuFe$_{1-x}$Al$_x$O$_2$ with $x=0.0155$. %
At first, we should review the magnetic phase transitions in the $x=0.0155$ sample.\cite{Terada_x_T} %
As shown in Fig. \ref{x-T-H_PhaseDiagram}, this system exhibits three magnetically ordered phases in zero field. %  %
For all the phases, the magnetic modulation wave vectors are described by $(0,q,\frac{1}{2}) [=(q_h,q_h,\frac{3}{2})_h]$. %
The highest temperature phase and the intermediate phase are the oblique partially disordered (OPD) phase %
and the partially disordered (PD) phase, respectively. %
Both of them have collinear sinusoidal magnetic structures with incommensurate wave numbers. %
The wave number of the OPD phase, $q^{\rm OPD}\sim0.390$, is independent of temperature, while %
that of the PD phase, $q^{\rm PD}$ varies with temperature ($0.404<q^{\rm PD}<0.430$). %
The lowest temperature phase is the FEIC phase, whose modulation wave number is $q^{\rm FEIC}=0.414$. % 

Figures \ref{BL46XU_results}(a)-(d) show the X-ray diffraction intensity maps around the reciprocal lattice position %
of $(2,2,0)_h$ in the paramagnetic (PM) phase and the magnetically ordered phases. %
As seen in Fig. \ref{BL46XU_results}(d), a single peak assigned as $(2,2,0)_h$ is observed  %
at $T=19.2$ K in the PM phase. %
In the PD and FEIC phases, this fundamental peak splits into several peaks, %
as shown in Figs. \ref{BL46XU_results}(a)-(b). %
This suggests that the threefold rotational symmetry along the $c_h$ axis vanishes, %
resulting in a monoclinic (or even lower symmetry) lattice distortion in the PD phase and the FEIC phase. %

The monoclinic lattice distortions are also %
observed in the undiluted ($x=0.00$) system,\cite{Terada_14.5T,Terada_CuFeO2_Xray} %
in which the PD and FEIC phases show up as the thermally induced phase and the first field-induced phase, %
respectively. %
As discussed in the previous studies,\cite{Terada_14.5T,Terada_CuFeO2_Xray,Ye_CuFeO2} these structural transitions are %
due to the bond order induced by the magnetostriction, which leads to lower-symmetry magnetic orderings and lifts the %
macroscopic degeneracy of the magnetic states. %

On the other hand, splitting of the $(2,2,0)_h$ peak was not observed in the OPD phase, as shown in Fig. \ref{BL46XU_results}(c), %
even though the magnetic structure of the OPD phase does not have threefold rotational symmetry along the $c_h$ axis.\cite{Terada_FONDER} %
This suggests that the coherent bond order is not essential to the OPD magnetic ordering. %
Based on the fact that the OPD phase is never observed without nonmagnetic substitution, it is reasonable to propose that the OPD %
magnetic ordering is stabilized by local symmetry breaking due to the site-random magnetic vacancies. %

Using the monoclinic basis, %
we can identify the splitting peaks in the FEIC and PD phases, as shown in Figs. \ref{BL46XU_results}(a) and \ref{BL46XU_results}(b), %
and we can also estimate the temperature variation of the lattice constants $a$ and $b$ in the $x=0.0155$ sample, %
as shown in Fig. \ref{BL46XU_results}(e). %
The differences between the lattice constants in the FEIC (or PD) phase and those in the PM phase in this system are %
comparable with those in the undiluted system.\cite{Terada_CuFeO2_Xray,Terada_14.5T,Ye_CuFeO2} %
These results suggest that the crystal structures of the FEIC and PD phases in the $x=0.0155$ sample %
are almost the same as those in the undiluted system. %

In the present measurements, %
we observed incommensurate superlattice reflection at the reciprocal lattice point of $(0,2.86,2)$ in the PD phase. % and the FEIC phase, %
This reflection can be assigned as $\mbox{\boldmath $\tau$}_{even}+(0,2q^{\rm PD},0)$, %
where $\mbox{\boldmath $\tau$}_{even}$ is the reciprocal lattice point of $(H,K,L)$ with the condition `$H+K=2n$' ($n$ is an integer). %
We also observed commensurate superlattice reflection at the reciprocal lattice points of $(0,3,2)$ in the FEIC phase.
The temperature variations of the intensities of these reflections are shown in Fig. \ref{BL46XU_results}(f). %
The incommensurate superlattice reflection in the PD phase and the commensurate superlattice reflection in the FEIC phase are also observed %
in the undiluted system,\cite{Terada_14.5T,Ye_CuFeO2,Terada_LatticeModulation} %
and consistently explained by the magnetostriction model proposed by Terada \textit{et al.}\cite{Terada_LatticeModulation} %
%The incommensurate superlattice reflection in the FEIC phase, that is the central issue of this paper, will be presented in the %
%following section. %

\begin{figure*}[t]
\begin{center}
	\includegraphics[keepaspectratio,clip,width=17cm]{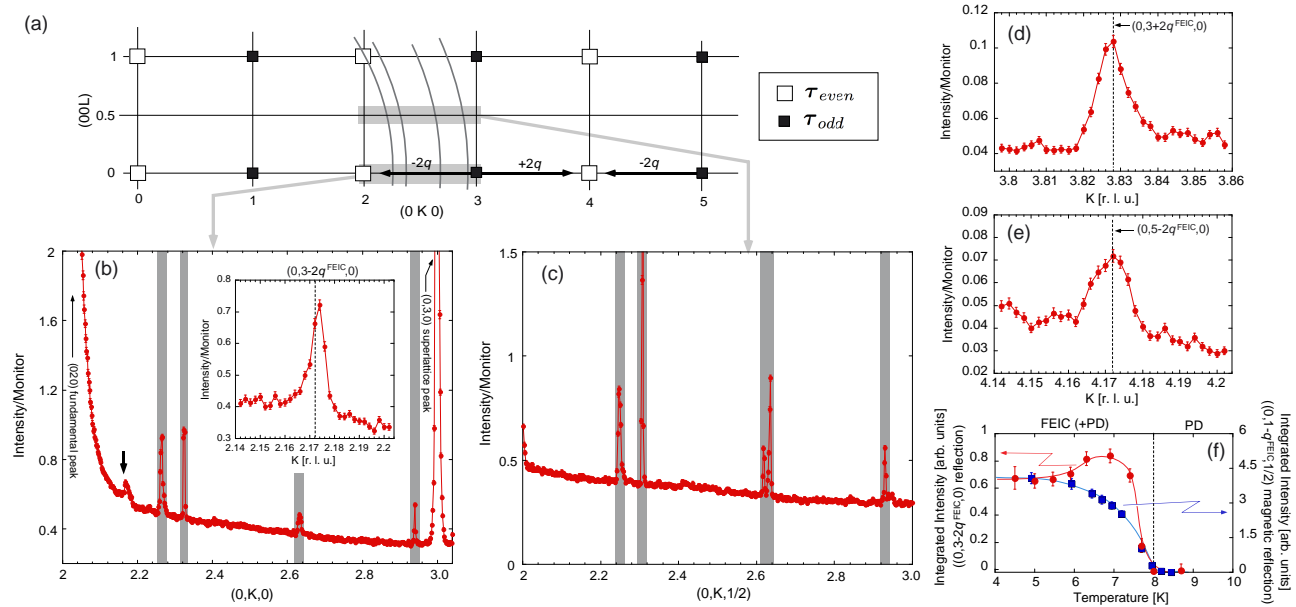}
	\caption{(Color online) (a) Illustration of the reciprocal lattice $(0KL)$ zone. %
	[(b)-(c)] Diffraction profiles of the (b) $(0K0)$ and (c)  $(0K\frac{1}{2})$ reciprocal lattice scans in the FEIC phase. %
	Contamination peaks from the Cu sample holder are masked by the shaded areas. %
	The inset of (b) shows the magnification around the reciprocal lattice position of $(0,3-2q^{\rm FEIC},0)$. %
	[(d)-(e)] The diffraction profiles of (d) $(0,3+2q^{\rm FEIC},0)$ and (e) $(0,5-2q^{\rm FEIC},0)$ reflections. %
	(f) Temperature dependences of the integrated intensity of the $(0,3-2q^{\rm FEIC},0)$ reflection and %
	the neutron diffraction intensity of the magnetic Bragg reflection corresponding to the FEIC magnetic ordering %
	(taken from Ref. \onlinecite{Terada_x_T}), under zero field. }%
	\label{BL22XU_LongProfile}
\end{center}
\end{figure*}

%These reflections are also observed in the undiluted system,\cite{Terada_14.5T,Ye_CuFeO2,Terada_LatticeModulation} %
%and consistently explained by the magnetostriction.\cite{Terada_LatticeModulation} %

\subsection{Lattice modulations in the FEIC phase}

We now focus on the incommensurate lattice modulations in the FEIC phase.
Figures \ref{BL22XU_LongProfile}(b) and \ref{BL22XU_LongProfile}(c) show the X-ray diffraction profiles of the %
$(0,K,0)$ and $(0,K,\frac{1}{2})$ reciprocal lattice scans at $T=4.5$ K in the FEIC phase under zero field. %
As shown in Fig. 4(b) and its inset, a satellite reflection is found at the reciprocal lattice position of (0, 2.172, 0). %
The intensity of this reflection is smaller than that of the $(0,4,0)$ fundamental reflection by a factor $10^{-7}\sim 10^{-8}$. %
This reflection is successfully identified as being $(0,3-2q^{\rm FEIC},0)$. %%
It should be noted that the experimental resolution in the present measurement can clearly distinguish the difference %
between the $q^{\rm FEIC}$ and $q^{\rm PD}$ phases at $T=4.5$ K $(\sim0.43)$, %
although the high-temperature PD phase coexists with the FEIC phase under zero field cooling in the $x=0.0155$ sample.\cite{Terada_x_T,CFAO_AnisoPD} %
We confirmed the repetition of this reflection in reciprocal lattice space, as shown in the Figs. \ref{BL22XU_LongProfile}(d) and \ref{BL22XU_LongProfile}(e). %
Hereafter, we refer to these reflections as $\mbox{\boldmath $\tau$}_{odd}\pm(0,2q^{\rm FEIC},0)$ reflections, %
where $\mbox{\boldmath $\tau$}_{odd}$ is the reciprocal lattice point of $(H,K,L)$ with the condition `$H+K=2n+1$'. %
%

%The temperature dependence of the intensity of the $(0,3-2q^{\rm FEIC},0)$ reflection is shown in Fig. \ref{BL22XU_LongProfile}(e). %
The $\mbox{\boldmath $\tau$}_{odd}\pm(0,2q^{\rm FEIC},0)$ reflections are observed only in the FEIC phase. %
However, as shown in Fig. \ref{BL22XU_LongProfile}(f), the temperature variation of the integrated intensity of the $\mbox{\boldmath $\tau$}_{odd}\pm(0,2q^{\rm FEIC},0)$ reflection is not %
proportional to that of the magnetic order parameter; %
specifically, the intensity of the $(0,3-2q^{\rm FEIC},0)$ reflection obviously increases around $T=6.5$ K with increasing temperature, %
and rapidly decreases at the transition temperature from the FEIC phase to the PD phase, $T=8$ K, %
while the FEIC magnetic order parameter (neutron diffraction intensity) monotonically decreases with increasing temperature, as seen in Fig. \ref{BL22XU_LongProfile}(f). %

No significant reflections were detected in the $(0,K,\frac{1}{2})$ reciprocal lattice scan, as shown in Fig. \ref{BL22XU_LongProfile}(c). %
The satellite reflections assigned as $(0,2n\pm q^{\rm FEIC},\frac{1}{2})$ %
and $(0,(2n+1)\pm q^{\rm FEIC},\frac{1}{2})$ were observed in the field-induced FEIC phase of undiluted CuFeO$_2$.\cite{Terada_LatticeModulation} %
These satellite reflections indicate the existence of a lattice modulation with a wave number $1q^{\rm FEIC}$ %
in the field-induced FEIC phase. %
However, these reflections were not observed in the zero-field FEIC phase of CuFe$_{1-x}$Al$_x$O$_2$ with $x=0.0155$. %
%feature of the FEIC phase, but the extrinsic feature induced by an application of a %
%magnetic field. %
This is consistent with the calculation of Terada \textit{et al.},\cite{Terada_LatticeModulation} which shows %
that the combination of the proper helical spin components and uniform magnetization components along the $c_h$ axis %
induces $1q$-lattice modulation through magnetostriction, %
and suggests that the $1q$-lattice modulation is not essential to the ferroelectricity in this system. %

The present results reveal that the $2q$-lattice modulation corresponding to the satellite reflections at %
$\mbox{\boldmath $\tau$}_{odd}\pm(0,2q^{\rm FEIC},0)$ exists in the zero-field FEIC phase. %
The microscopic origin of the $2q$-lattice modulation is discussed in sec. IV. %

\subsection{$H_{\parallel b}$ dependence of the $2q$-lattice modulation}

\begin{figure}[t]
\begin{center}
	\includegraphics[keepaspectratio,clip,width=7.8cm]{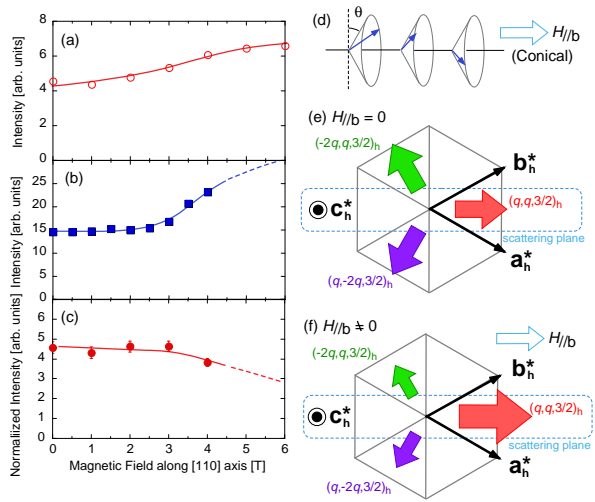}
	\caption{(Color online) (a)The  $H_{\parallel b}$-dependence of the integrated intensity of $(0,3-2q^{\rm FEIC},0)$ reflection observed in the X-ray diffraction measurements. %
	(b) The $H_{\parallel b}$-dependence of $(\frac{1}{2}-q_h^{\rm FEIC},\frac{1}{2}-q_h^{\rm FEIC},\frac{3}{2})_h$ magnetic Bragg reflection observed in the neutron diffraction measurements at $T=4.5$ K. %
	(d) The field dependence of the integrated intensity of $(0,3-2q^{\rm FEIC},0)$ reflection normalized by the data of the neutron %
	diffraction measurements. %
	(d) Schematic drawings of a conical structure with uniform magnetization %
	component along the helical axis. %
	[(e)-(f)] Schematic drawings of the fractions of three magnetic domains (e) under zero field, and (f) under applied field along %
	the $b$ axis ($[110]_h$ axis). The directions of the filled arrows denote the $[001]_h$ projections of the propagation wave vectors. %
	The sizes of arrows qualitatively show the fractions of each domain. }
	\label{H110_dependence}
\end{center}
\end{figure}

We also surveyed the magnetic field dependence of the $\mbox{\boldmath $\tau$}_{odd}\pm(0,2q^{\rm FEIC},0)$ reflections by %
the X-ray diffraction measurements under an applied field. %
As mentioned above, application of a magnetic field along the $c_h$ axis readily leads to %
additional lattice modulations through magnetostriction. %
We thus applied a magnetic field along the helical axis of the proper helical magnetic structure, i.e. along the $b$ axis. %
When a magnetic field is applied along this direction, the proper helical magnetic structure %
deforms into a conical magnetic structure, %
as shown in Fig. \ref{H110_dependence}(d). %
In this configuration, the magnetic field ($H_{\parallel b}$) does not modify the amplitude of the magnetic moment at each Fe$^{3+}$ site. %
Hence, the magnetostriction cannot induce additional lattice modulations. %

Figure \ref{H110_dependence}(a) shows the integrated intensity of the $(0,3-2q^{\rm FEIC},0)$ reflection at $T=4.5$ K as a function of %
the magnetic field along the $b$ axis ($H_{\parallel b}$). %
The intensity of the reflection increases with increasing magnetic field. %
This result leads to the following two possibilities: %
that the amplitude of the $2q$-lattice modulation itself increases with increasing magnetic field, or %
the fractions of the three magnetic domains, which reflect the threefold rotational symmetry of the original trigonal crystal structure, %
change with magnetic field intensity. %

We have also performed $(0,K,0)$ and $(0,K,\frac{1}{2})$ reciprocal lattice scans at $T=4.5$ K under $H_{\parallel b}=6$ T. %
In both these scans, no additional (field-induced) reflections are detected. %

\subsection{Neutron diffraction measurements under applied field along the $b$ axis}%
In order to elucidate the origin of the $H_{\parallel b}$-variation of the intensity of the $(0,3-2q^{\rm FEIC},0)$ reflection, %
we performed neutron diffraction measurements under an applied field along the $b$ axis. % 
We found that the magnetic Bragg reflections corresponding to the FEIC %
magnetic ordering increase with increasing magnetic field.\cite{Comment1} %
Figure \ref{H110_dependence}(b) shows the field dependence of the intensity of the $(\frac{1}{2}-q_h^{\rm FEIC},\frac{1}{2}-q_h^{\rm FEIC},\frac{3}{2})_h$ magnetic %
Bragg reflection, where $q_h^{\rm FEIC}[=\frac{1}{2}q^{\rm FEIC}]=0.207$. %
Since the change in the magnetic structure factor is considered to be negligible for magnetic fields of $0<H_{\parallel b}<4$ T,\cite{CFAO_AnisoPD} %
this enhancement of the magnetic Bragg intensity indicates that the fraction of the FEIC magnetic domain increases with the propagation %
wave vector $(q_h^{\rm FEIC},q_h^{\rm FEIC},\frac{3}{2})_h$. % 
This result supports the latter scenario for the enhancement of the intensity of the $(0,3-2q^{\rm FEIC},0)$ reflection. %, 
A magnetic field along the $b$ axis ([110]$_h$ axis) should favor the magnetic domain in which a more uniform magnetization component %
is induced along the direction of the magnetic field. %
As a result, the fraction of the magnetic domains with a wave vector $(q_h,q_h,\frac{3}{2})_h$ is enhanced %
by the magnetic field along the $[110]_h$ axis, and the fractions of the other magnetic domains with the wave vectors $(-2q_h,q_h,\frac{3}{2})_h$ and $(q_h,-2q_h,\frac{3}{2})_h$ %
should be reduced, as shown in Figs. \ref{H110_dependence}(e) and \ref{H110_dependence}(f). %
%
%We confirmed that this $H_b$-variation of the FEIC magnetic Bragg reflections is also observed in the single FEIC phase, which is prepared by the %
%cooling process under applied field along the $[001]_h$ axis. %
%This indicates that the existence of the PD phase, which coexists with the FEIC phase under zero field, is not relevant to the enhancement of the %
%FEIC magnetic domain with the wave vector $(q_h,q_h,\frac{3}{2})_h$. %
%
The details of the magnetic domain distributions and the magnetic phase transitions in CuFe$_{1-x}$Al$_x$O$_2$ ($x=0.0155$) %
under applied field along the $[110]_h$, $[1\bar{1}0]_h$ and $[001]_h$ directions will be presented in another paper.\cite{CFAO_AnisoPD} %

Figure \ref{H110_dependence}(c) shows the intensity of the $(0,3-2q^{\rm FEIC},0)$ reflection normalized to the intensity of the magnetic Bragg reflection. %
In the magnetic field region of $H_{\parallel b}\le4$ T, the amplitude of the $2q$-lattice modulation does not exhibit remarkable change, %
and it decreases slightly above $H_{\parallel b}=3$ T. %

\begin{figure}[b]
\begin{center}
	\includegraphics[keepaspectratio,clip,width=7.8cm]{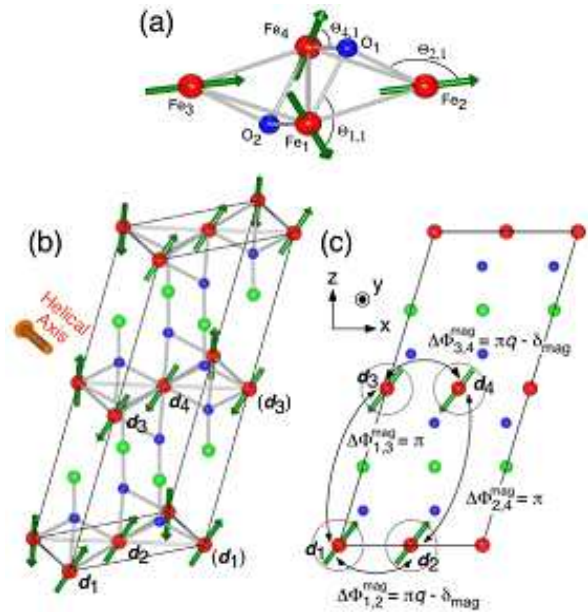}
	\caption{(Color online) (a) A Fe$_4$O$_2$ cluster used for the calculation of the local electric dipole moment in Ref.\onlinecite{Arima_Symmetry}. %
	(b) The magnetic structure in the FEIC phase with the $a\times b\times 2c$ cell. %
	(c) The phase differences in the magnetic modulations ($\Delta \Phi_{i,j}^{mag}$) between the Fe$^{3+}$ sites. }
	\label{unit_cell}
\end{center}
\end{figure}

\section{DISCUSSIONS}

\subsection{Review of Arima's cluster model}
In this section, we discuss whether the microscopic origin of the $2q$-lattice modulation observed in the FEIC phase %
is the variation in the $d$-$p$ hybridization or not. %
At first, we start from the Fe$_4$O$_2$ cluster model used in Ref. \onlinecite{Arima_Symmetry} (see Fig. \ref{unit_cell}(a)). %
Applying the microscopic theory by Jia \textit{et al.}\cite{Jia1,Jia2} to the Fe-O covalent bonding, %
Arima pointed out that the $d$-$p$ hybridization between a magnetic Fe$^{3+}$ ion and a ligand O$^{2-}$ ion is %
slightly modified depending on the direction of the magnetic moment at the Fe$^{3+}$ site.\cite{Arima_Symmetry} %c
As a result, the charge transfer from the magnetic Fe$_i$ ($i=1,2,3,4$) site to the neighboring O$_j$ ($j=1,2$) site %
can be described by %
\begin{eqnarray}
C_0 + \Delta C\cos2\Theta_{i,j},
\label{2C_eq}
\end{eqnarray}
where $C_0$ is the charge transfer in the PM phase, and $\Delta C$ %
is the constant dependant on the difference in the spin-orbit interaction between Fe$_i$ and O$_j$, and %
the magnitude of the ordered magnetic moment at the Fe site, and so on. %
$\Theta_{i,j}$ is the angle between the direction of the Fe$_i$-O$_j$ bond and the direction of the magnetic moment at the Fe$_i$ site. %
The macroscopic uniform electric polarization is not induced by this term only. %
Arima argued that the many-body effect among the three covalent Fe-O bonds which are concerned with an O$^{2-}$ ion, should be %
taken into account in Eq. (\ref{2C_eq}). %
As shown in Fig. \ref{unit_cell}(a), an O$^{2-}$ ion in this cluster (or CuFeO$_2$) is surrounded by three neighboring Fe$^{3+}$ ions. %
Hence, the charge transfer between Fe$_i$ and O$_j$ should be slightly affected by the amount of charge transfer in the other two Fe-O bonds %
which are concerned with the O$_j$ site. %
For example, the covalency between O$_1$ and Fe$_1$ should be reduced when Fe$_2$-O$_1$ and %
Fe$_4$-O$_1$ bonds are more covalent. %
Taking account of this many-body effect, the charge transfer from Fe$_1$ to O$_1$ can be described as follows:\cite{Comment2} %
\begin{eqnarray}
C_0[1+C'\cos2\Theta_{1,1}][1-\alpha C'(\cos2\Theta_{2,1}+\cos2\Theta_{4,1})],
\label{Eq_CT}
\end{eqnarray}
where $C'=\frac{\Delta C}{C_0}(\ll1)$. $\alpha$ is the parameter representing the efficiency of the many-body effect, and is assumed to be small %
$\alpha \ll 1$. %
Here, we apply to the cluster, a proper helical magnetic structure with a modulation wave number $q$, whose helical axis is parallel to the Fe$_1$-Fe$_4$ bond direction. %
The helical-axis component of the induced electric dipole moment at the O$_1$ site, $p_b$, %
should be proportional to the imbalance between the charge transfer from Fe$_1$ to O$_1$ and that from Fe$_4$ to O$_1$, %
\begin{eqnarray}
p_b&\propto& (1+\alpha)(\cos2\Theta_{1,1}-\cos2\Theta_{4,1})\nonumber\\
&&-\alpha C'\cos2\Theta_{2,1}(\cos2\Theta_{1,1}-\cos2\Theta_{4,1}).
\label{Eq_Pb}
\end{eqnarray}
In the case of the proper helical magnetic structure, the value of $\cos2\Theta_{i,j}$ oscillates with a period of half the magnetic modulation. %
As a result, the first term of the Eq. (\ref{Eq_Pb}) gives the nonuniform polarization oscillating with a wave number of $2q$ ($2q$-modulation), %
and the second term of Eq. (\ref{Eq_Pb}) gives the uniform polarization and the nonuniform polarization oscillating with a wave number of $4q$ %
($4q$-modulation). %
It should be noted that one can derive Eq. (6) in Ref. \onlinecite{Arima_Symmetry} from Eq.(\ref{Eq_Pb}) in this paper. %
The above formula of the local electric polarization suggests that the amplitude of the $4q$-modulation is much smaller than that of the $2q$-modulation. %
In the limit of $\alpha = 0$, the uniform polarization and $4q$-modulation vanish. %
It is worth mentioning here that the same amount of electric dipole moment is induced at the O$_2$ site, %
because of the symmetry of the proper helical magnetic structure. % 

\begin{figure}[b]
\begin{center}
	\includegraphics[keepaspectratio,clip,width=7.8cm]{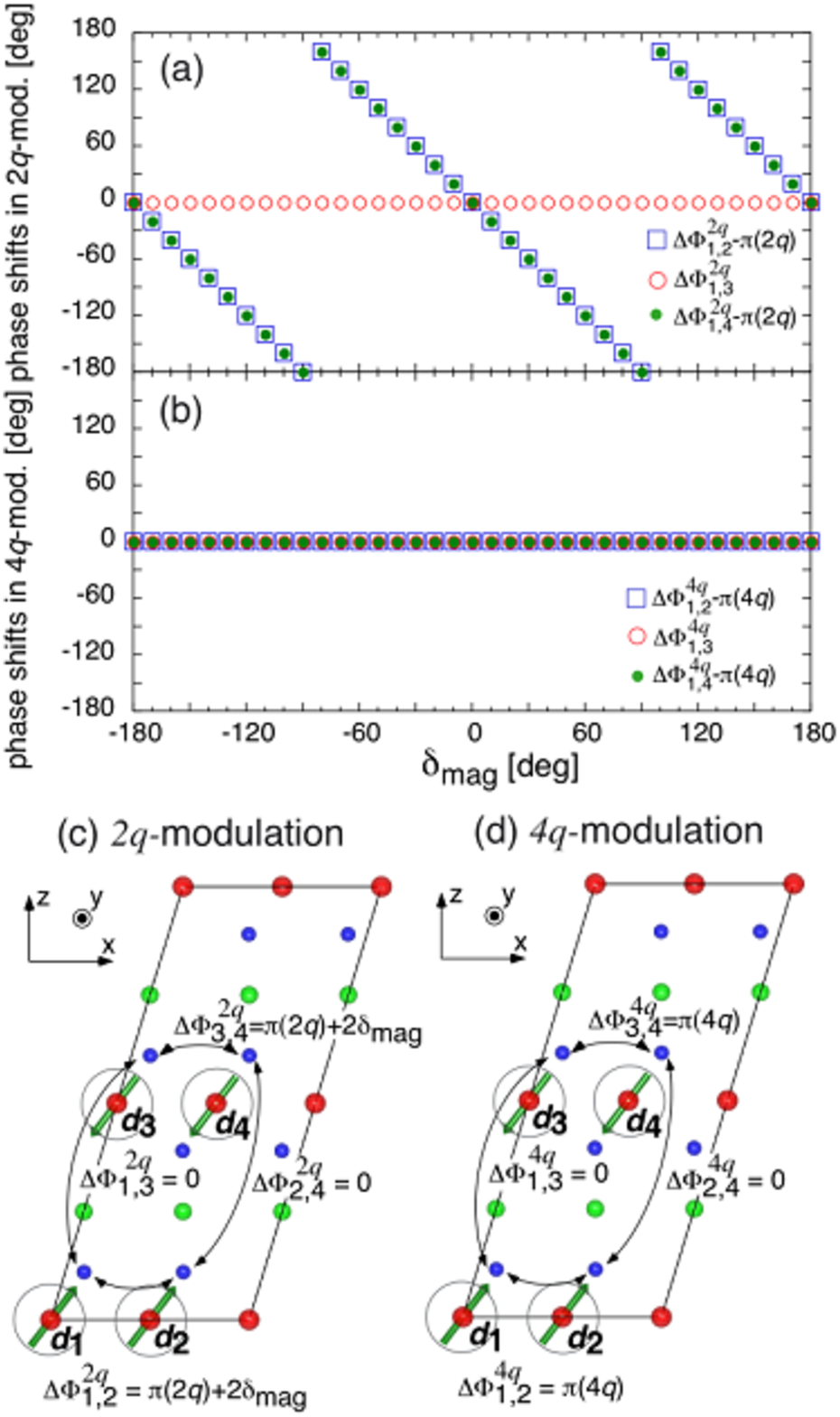}
	\caption{(Color online) [(a)-(b)] $\delta_{mag}$ dependence of the difference between the phases of (a) $2q$- and (b) $4q$-modulations. %
	[(c)-(d)] The phase differences of the (c) $2q$- and (d) $4q$-components in the spatial modulations in the local electric polarization. }
	\label{Fourier_calc}
\end{center}
\end{figure}

\subsection{Calculation results in zero field}
We apply Eq. (\ref{Eq_Pb}) to the magnetic structure in the FEIC phase, which has been determined by a previous magnetic structure analysis.\cite{SpinNoncollinearlity} %
For simplicity, hereafter, we refer to the wave number and the wave vector in the FEIC phase as $q$ and $\mbox{\boldmath $q$}[=(0,q,\frac{1}{2})]$, %
respectively. %
In order to define the spin arrangement in the FEIC phase, we employ $a\times b\times 2c$ cell, as shown in Fig. \ref{unit_cell}(b). %
%The schematic drawings of the magnetic unit cell and the magnetic structure in the FEIC phase are shown in Figs. \ref{unit_cell}(b) and \ref{unit_cell}(c). %
This cell contains four Fe$^{3+}$ sites, and the fractional coordinates of these sites are:
\begin{eqnarray}
\mbox{\boldmath $d$}_1 &=& (0,0,0)\nonumber\\
\mbox{\boldmath $d$}_2 &=& (0,1/2,0)\nonumber\\
\mbox{\boldmath $d$}_3 &=& (0,0,1)\nonumber\\
\mbox{\boldmath $d$}_4 &=& (0,1/2,1).
\end{eqnarray}
From the results of a previous magnetic structure analysis, the spin components at $\mbox{\boldmath $d$}_i$ site are described, %
using the Cartesian coordinates shown in Fig. \ref{unit_cell}(c), as follows:
\begin{eqnarray}
S_i^x &=& \mu_x \cos(2\pi\mbox{\boldmath $q$}\cdot(\mbox{\boldmath $l$}+\mbox{\boldmath $d$}_i)  - \phi_i)\nonumber\\
S_i^y &=& 0 \nonumber\\
S_i^z &=& \mu_z \sin(2\pi\mbox{\boldmath $q$}\cdot(\mbox{\boldmath $l$}+\mbox{\boldmath $d$}_i)  - \phi_i)
\label{S_component}
\end{eqnarray}
$\mu_x$ and $\mu_z$ are the magnetic moments along the $x$ and $z$ axes, respectively. %
Since no significant ellipticity has been detected in the magnetic structure analysis for zero field (and for relatively low fields, $H_{\parallel[001]_h}<4$T) on %
the $x=0.0155$ sample,\cite{SpinNoncollinearlity} we assume at this stage that the magnetic structure has no ellipticity, i.e., $\mu_x = \mu_z=\mu$. %
$\phi_i$ is the relative phase shift at the $\mbox{\boldmath $d$}_i$ site; specifically, % 
\begin{eqnarray}
|\phi_1-\phi_2|&=&|\phi_3-\phi_4|=\delta_{mag}\nonumber\\
|\phi_1-\phi_3|&=&|\phi_2-\phi_4|=0
\end{eqnarray}%
Although $\delta_{mag}$ was determined to be $\sim76^{\circ} (\sim \pi q)$ by magnetic structure analysis,\cite{SpinNoncollinearlity} %
we treat $\delta_{mag}$ as a parameter in the following calculation. % 
Here, we define the phase difference of the magnetic modulations between the $\mbox{\boldmath $d$}_i$ site and the $\mbox{\boldmath $d$}_j$ site as $\Delta\Phi_{i,j}^{mag}$. %
Using this notation, the relationship among the phases of the magnetic modulations on the four Fe$^{3+}$ sites is %
summarized in Fig. \ref{unit_cell}(c). %

\begin{figure}[t]
\begin{center}
	\includegraphics[keepaspectratio,clip,width=6.9cm]{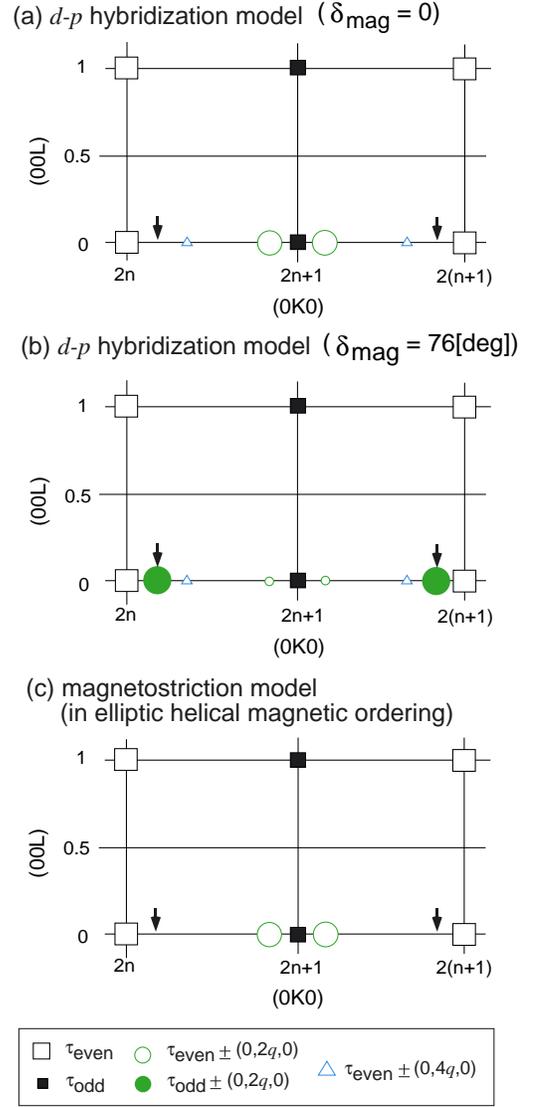}
	\caption{(Color online) Fourier spectrum mappings of the spatial modulations in the local polarization, which are calculated using %
	(a) the $d$-$p$ hybridization model with $\delta_{mag}=0,\pi$, %
	(b) that with $\delta_{mag}=76^{\circ}$, and %
	(c) the magnetostriction model with finite ellipticity. %
	The size of a symbol, except for $\mbox{\boldmath $\tau$}_{even}$ and $\mbox{\boldmath $\tau$}_{odd}$, qualitatively shows the intensity of the spectrum $|P(\mbox{\boldmath $Q$})|^2$, %
	which is normalized by the intensity of the most strongest peak in the spectrum. % 	
	The black arrows denote the position of the satellite reflections observed in the zero-field FEIC phase. }
	\label{Fourier_map}
\end{center}
\end{figure}

We numerically calculated the spatial modulations of $p_b$ at each oxygen site using Eq. (\ref{Eq_Pb}). %
The system size is set to be $a\times Nb\times2c$. %
Since the magnetic propagation wave number along the $b$ axis is incommensurate, $N$ is set to be a large number (typically, $N\sim100$). %
We also calculated Fourier spectra for their nonuniform components. %
Here, we define the phase difference in the $nq$-modulation ($n=2,4$) between the oxygen site neighboring $\mbox{\boldmath $d$}_i$ site and that neighboring $\mbox{\boldmath $d$}_j$ site, %
as $\Delta\Phi_{i,j}^{nq}$, in the same manner as that of the magnetic modulations. %
The results of the calculations for $\Delta\Phi_{i,j}^{2q}$ and $\Delta\Phi_{i,j}^{4q}$ are shown in Figs. \ref{Fourier_calc}(a) and \ref{Fourier_calc}(d). %
For the $2q$-modulations, there is a finite phase shift, $2\delta_{mag}$, between the oxygen %
site neighboring the $\mbox{\boldmath $d$}_1$ ($\mbox{\boldmath $d$}_3$) %
site and that neighboring the $\mbox{\boldmath $d$}_2$ ($\mbox{\boldmath $d$}_4$) site. %
For the $4q$-modulation, there are no phase shifts along the $a$ axis.  %
For both of the $2q$- and $4q$-modulations, there are also no phase shifts along the $c$ axis. %
These results are summarized in Figs. \ref{Fourier_calc}(c) and \ref{Fourier_calc}(d). %

The intensities of the calculated Fourier spectra for the spatial modulations of $p_b$, $|P(\mbox{\boldmath $Q$})|^2$, %
where $\mbox{\boldmath $Q$}$ is a vector in the reciprocal lattice space, are mapped onto the reciprocal lattice space, %
as shown in Figs. \ref{Fourier_map}(a) and \ref{Fourier_map}(b). %
Since the $d$-$p$ hybridization model predicts that the local electric polarization arises from the local imbalance of the Fe-O bond-covalency, %
it is reasonable to consider that the spatial modulations of the local electric polarization result in spatial modulations of %
the local atomic displacements of the oxygen ions. %
We thus assume that the local displacement of the oxygen ion is proportional to the magnitude of the local polarization vector. %
In this case, $|P(\mbox{\boldmath $Q$})|$ is proportional to the structure factor in the X-ray diffraction measurements. %
In the case of $\delta_{mag}=0,\pi$, satellite peaks corresponding to the $2q$- and $4q$-modulations %
appear at $\mbox{\boldmath $\tau$}_{even}\pm(0,2q,0)$ and $\tau_{even}\pm(0,4q,0)$, respectively.
In the case of $\delta_{mag}\ne0,\pi$, satellite peaks corresponding to the $2q$-modulation appear, % 
in addition to the above peaks, at $\mbox{\boldmath $\tau$}_{odd}\pm(0,2q,0)$.
By substituting $\delta_{mag}=76^{\circ}$, the ratio of the intensities of the spectra, $|P(\mbox{\boldmath $\tau$}_{even}\pm(0,2q,0))|^2$ %
and $|P(\mbox{\boldmath $\tau$}_{odd}\pm(0,2q,0))|^2$ is calculated to be %
\begin{eqnarray}
\frac{|P(\mbox{\boldmath $\tau$}_{even}\pm(0,2q,0))|^2}{|P(\mbox{\boldmath $\tau$}_{odd}\pm(0,2q,0))|^2}=0.062.
\end{eqnarray}
This suggests that in X-ray measurements, the satellite reflections corresponding to the $2q$-modulation should be mainly observed at %
the reciprocal lattice position of $\mbox{\boldmath $\tau$}_{odd}\pm(0,2q,0)$. %
This shows good agreement with the results of the present measurements. %
Although reflections at $\mbox{\boldmath $\tau$}_{even}\pm(0,2q,0)$ and $\mbox{\boldmath $\tau$}_{even}\pm(0,4q,0)$ were not observed in the present measurements, %
this can be ascribed to the S/N of the present measurements, because the intensities of those reflections are %
estimated to be much smaller than that of $\mbox{\boldmath $\tau$}_{odd}\pm(0,2q,0)$ reflections. %

It should be noted that this result does not change in the parameter region of $0<\alpha<1$, %
although we assumed $\alpha = 0.1$ in the above calculation. %
This is because the value of $\alpha$ affects only the ratio between the amplitude of the $2q$-modulation and that of the $4q$-modulation (and uniform polarization); %
specifically, $|P(\mbox{\boldmath $\tau$}_{even}\pm(0,4q,0))|^2/|P(\mbox{\boldmath $\tau$}_{odd}\pm(0,2q,0))|^2$ %
is in the order of $\sim\alpha^2 C'^2 (\ll1)$

It is worth mentioning here that neither the magnetostriction model nor the spin-current model explains %
the satellite reflections at $\mbox{\boldmath $\tau$}_{odd}\pm(0,2q,0)$, even if the magnetic structure has finite ellipticity. %
The spin-current term, $\mbox{\boldmath $e$}_{i,i+1}\times (\mbox{\boldmath $S$}_i \times \mbox{\boldmath $S$}_{i+1})$, %
does not produce the helical-axis component of the local polarization vector, in the proper (or elliptic) helical magnetic ordering in this system. %
Therefore, the lattice modulations induced by this term do not contribute to the satellite reflection in the reciprocal lattice position of $(0,K,0)$. %
The magnetostriction term, ($\mbox{\boldmath $S$}_i\cdot \mbox{\boldmath $S$}_{i+1}$), produces the spatial modulation with a wave number of $2q$ %
in the helical-axis components of the local polarization vectors, when the magnetic structure has finite ellipticity.\cite{Terada_LatticeModulation} %
However, as discussed by Terada \textit{et al.},\cite{Terada_LatticeModulation} %
the satellite reflections corresponding to the magnetostriction-induced $2q$-modulation are observed only at %
the reciprocal lattice position of $\mbox{\boldmath $\tau$}_{even}\pm(0,2q,0)$, regardless of the value of $\delta_{mag}$ (see Fig. \ref{Fourier_map}(c)). %
Therefore, we conclude that the origin of the $2q$-lattice modulation observed in the zero-field FEIC phase is the %
variation in the metal-ligand hybridization with spin-orbit coupling, which corresponds to the feature (iii) mentioned in the introduction. %

We should mention the non-monotonic temperature variation of the intensity of the $\mbox{\boldmath $\tau$}_{odd}\pm(0,2q,0)$ reflection observed in the present measurement (see Fig. \ref{BL22XU_LongProfile}(f)). %
The $d$-$p$ hybridization between the Fe$^{3+}$ ions and the O$^{2-}$ ions must depend on the length of the %
Fe-O bonds as well as the angle between the Fe-O bond direction and the magnetic moment at the Fe site. %
We therefore anticipate that the temperature variations of the lattice constants and the fractional coordinates of the %
oxygen sites affect the temperature variation of the intensity of the $\mbox{\boldmath $\tau$}_{odd}\pm(0,2q,0)$ reflection. %
The precise determination of the structural parameters by high-resolution X-ray %
and/or neutron diffraction measurements are desirable for further clarification. %

\begin{figure}[t]
\begin{center}
	\includegraphics[clip,keepaspectratio,width=7.5cm]{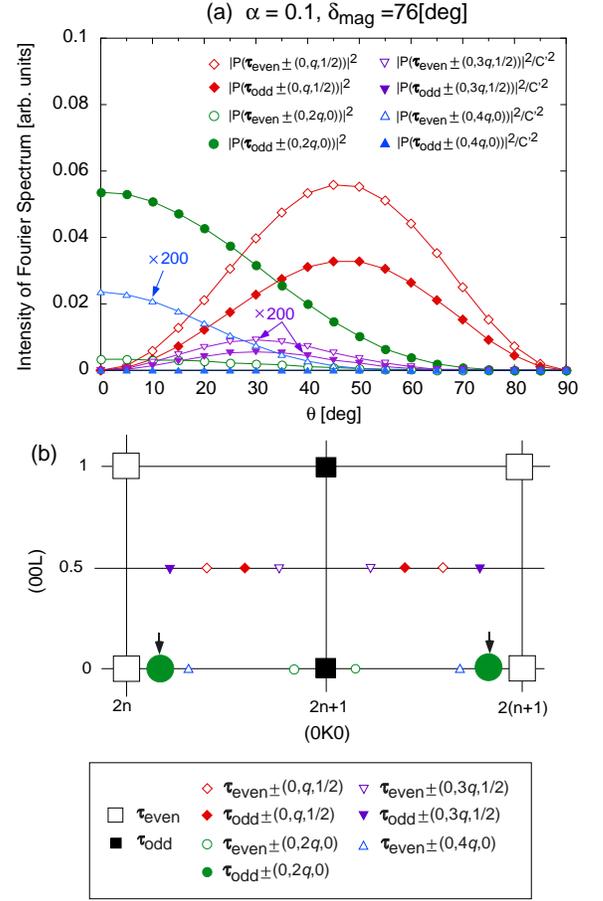}
	\caption{(Color online) (a) $\theta$ dependence of the intensities of the Fourier spectra for the spatial modulations in the local polarization. %
	To enhance the visibility, the data of $3q$ and $4q$ are magnified by a factor of 200. % 
	(b) Fourier spectrum mapping of the spatial modulations in the local polarization, which are calculated with %
	the $d$-$p$ hybridization model with $\delta_{mag}=76^{\circ}$, $\alpha=0.1$ and $\theta=10^{\circ}$. %
	The size of the symbols, except for $\mbox{\boldmath $\tau$}_{even}$ and $\mbox{\boldmath $\tau$}_{odd}$, corresponds to the intensity of the spectrum $|P(\mbox{\boldmath $Q$})|^2$, %
	which are normalized using the intensity of the most intense peak in the spectrum. % 
	The black arrows denote the positions of the satellite reflections observed in the present measurements. }
	\label{H110_calc}
\end{center}
\end{figure}

\subsection{Calculation results for an applied field }

When a magnetic field is applied along the $b$ axis, the magnetic structure is expected to be a conical structure. %
We also numerically calculated the Fourier spectra for the nonuniform polarizations induced by the conical magnetic structure. %
Figures \ref{H110_calc}(a) and \ref{H110_calc}(b) show the calculation results with the parameters of $\alpha =0.1$ and $\delta_{mag}=76^{\circ}$. %
$\theta$ is the angle defined in the Fig. \ref{H110_dependence}(a). %
With increasing $\theta$, the intensities of the Fourier spectra for 2$q$- and 4$q$-modulations decrease. %
In the region of $0<\theta<\frac{\pi}{2}$, $1q$- and $3q$-modulations emerge from %
the first and second terms of Eq. (\ref{Eq_Pb}), respectively. %
In reciprocal lattice space, the satellite peaks corresponding to the $1q$-modulations appear at %
$\mbox{\boldmath $\tau$}_{even}\pm(0,q,\frac{1}{2})$ and $\mbox{\boldmath $\tau$}_{odd}\pm(0,q,\frac{1}{2})$, and the satellite peaks corresponding to the $3q$-modulations appear at %
$\mbox{\boldmath $\tau$}_{even}\pm(0,3q,\frac{1}{2})$ and $\mbox{\boldmath $\tau$}_{odd}\pm(0,3q,\frac{1}{2})$, as shown in Fig. \ref{H110_calc}(b). %

According to the results of recent magnetization measurements,\cite{PulsMag_CFAO} the canting angle $\theta$ at $H_{\parallel b}=4$T %
is roughly estimated to be $5^{\circ}\sim10^{\circ}$. %
In the region of $\theta<10^{\circ}$, %
$|P(\mbox{\boldmath $\tau$}_{odd}\pm(0,2q,0))|^2$ decreases slightly with increasing $\theta$, and the intensities of the other Fourier spectra are quite small %
compared with $|P(\mbox{\boldmath $\tau$}_{odd}\pm(0,2q,0))|^2$, as shown in Fig. \ref{H110_calc}(a). %
Hence, it is expected that the intensity of the $\mbox{\boldmath $\tau$}_{odd}\pm (0,2q,0)$ satellite reflection %
decreases slightly with increasing applied magnetic field along the $b$ axis. %
This is consistent with the results of the present measurements. %
Although we could not survey the region of $\theta>10^{\circ}$, where remarkable changes in the amplitudes of the lattice modulations %
are expected, %
these results imply that the $d$-$p$ hybridization model still works for a finite magnetic field along the $b$ axis. %

\section{SUMMARY}
In summary, we have performed synchrotron radiation X-ray and neutron diffraction measurements under zero field and %
applied field, on the delafossite multiferroic CuFe$_{1-x}$Al$_x$O$_2$ %
with $x=0.0155$, in which the ferroelectric phase shows up under zero field. %

We found that the threefold rotational symmetry along $c_h$ axis vanishes in the PD and FEIC phases. %
%a symmetry-lowering structural transition corresponding to the magnetic transition from the OPD phase %
%to the PD phase. %
Although the present result does not determine the symmetry of the crystal structure in the PD and FEIC phases, %
it is reasonable to propose that a monoclinic lattice distortion occurs in the FEIC and PD phases in CuFe$_{1-x}$Al$_x$O$_2$ with $x=0.0155$, % 
because the monoclinic lattice distortion is observed in the PD and FEIC phases in undiluted CuFeO$_2$.\cite{Terada_CuFeO2_Xray,Terada_14.5T,Ye_CuFeO2} %
On the other hand, monoclinic lattice distortion was not observed in the OPD phase, which never shows up %
without nonmagnetic substitution. %
This implies that the OPD magnetic ordering is stabilized not by the coherent bond order due to magnetostriction, %
but by the local symmetry breaking due to site-random magnetic vacancies. %

In the FEIC phase, we found satellite reflections identified as $\mbox{\boldmath $\tau$}_{odd}\pm (0,2q^{\rm FEIC},0)$ under zero field. %
This indicates that the FEIC phase is essentially accompanied by $2q$-lattice modulations. %
We have calculated the Fourier spectra of the spatial modulations in the local electric polarization using the experimentally determined magnetic structure in the FEIC phase and %
the microscopic model presented by Arima.\cite{Arima_Symmetry} %
Comparing the experimental results with the calculated Fourier spectrum revealed that the origin of the $2q$-lattice modulation is not %
conventional magnetostriction but the variation in the $d$-$p$ hybridization between the magnetic Fe$^{3+}$ ions and the ligand O$^{2-}$ ions. %
Combining the present results with the results of a previous polarized neutron diffraction study,\cite{CFAO_Helicity} %
we conclude that the microscopic origin of the ferroelectricity in this system is the variation in the $d$-$p$ hybridization %
with spin-orbit coupling. %

\section*{Acknowledgment}

The authors are grateful to T. Arima and S. Onoda for helpful discussions about the mechanism of the microscopic spin-polarization coupling, %  
and are also grateful to N. Nagaosa for suggesting the importance of the lattice modulations in a new class of magnetoelectric multiferroics. %  
The neutron diffraction measurement at BENSC was carried out along the proposal PHY-01-2285-DT. %
The synchrotron radiation X-ray diffraction measurements at SPring-8 were performed along the proposal %
2007B3773 (BL22XU) with the approval of the nano-technology network project, and the proposal 2006B1348 (BL46XU). %
This work was supported by a Grant-in-Aid for Scientific Research (C), No. 19540377, from JSPS, Japan. %
The images of the crystal and magnetic structures in this paper were depicted using the software VESTA\cite{VESTA}  %
developed by K. Monma.%


\begin{thebibliography}{99}
\bibitem{Comment1} In both of the X-ray and neutron diffraction measurements, $H_{\parallel b}$-increasing scans were performed after zero field cooling, in which the high-temperature PD phase coexists with the FEIC phase. %
However, we should mention that the same $H_{\parallel b}$-dependence of the FEIC magnetic Bragg reflection is observed in the %
single FEIC phase, which is prepared by the cooling process with an applied field along $[001]_h$ axis. %
\bibitem{Comment2} The many-body effect can be introduced as $C_0(1+C'\cos2\Theta_{1,1})(1-\alpha C'\cos2\Theta_{2,1})(1-\alpha C'\cos2\Theta_{4,1})$. Here, we neglect the term of $O(\alpha^2)$, and obtain eq. (\ref{Eq_CT}). %
\end{thebibliography}
\end{document}